\documentclass[12pt]{article}
\usepackage{amsmath,amssymb,amsfonts,epsfig,graphicx}
\usepackage{amsmath,amssymb,epsfig}
\numberwithin{equation}{section}
\newcommand{\be}{\begin{equation}}
\newcommand{\bea}{\begin{eqnarray}}
\newcommand{\eea}{\end{eqnarray}}
\newcommand{\ba}{\begin{array}}
\newcommand{\ea}{\end{array}}
\newcommand{\ee}{\end{equation}}

\expandafter\ifx\csname mathbbm\endcsname\relax

\else

\newcommand{\bse}{\begin{subequations}}
\newcommand{\ese}{\end{subequations}}
\fi \textheight 22cm \textwidth 15cm \topmargin 1mm \oddsidemargin
5mm \evensidemargin 5mm

\def\by{\times}

\def\N{{\cal N}}

\def\opt{operator\ }

\def\susy{supersymmetry\ }

\def\sugra{supergravity\ }

\def\super{$PSU(2|2)\times PSU(2|2)\times U(1)$}
\def\ads{$AdS_5\times S^5$}
\def\psutwo{$PSU(2,2|4)$}

\begin{document}

\begin{titlepage}
 \hfill
 \vbox{
    \halign{#\hfil         \cr
           IPM/P-2005/070 \cr
           } 
      }  
 \vspace*{20mm}
 \begin{center}
 {\Large {\bf Semi-classical Probe Strings on Giant Gravitons Backgrounds}\\ }

 \vspace*{15mm} \vspace*{1mm} {Mohsen Alishahiha$^1$, Hajar Ebrahim$^1$, Batool Safarzadeh$^2$ and \\
 Mohammad M. Sheikh-Jabbari$^1$}

 \vspace*{1cm}

${}^1$ {\it  Institute for Studies in Theoretical
 Physics and Mathematics (IPM)\\
 P.O. Box 19395-5531, Tehran, Iran\\}

\vspace*{.4cm}

${}^2$ {\it Department of physics, School of Science\\
Tarbiat Modares University, P.O. Box 14155-4838, Tehran, Iran}

 \vspace*{1cm}
 \end{center}

 \begin{abstract}
 In the first part of this paper we study two $\mathbb{Z}_2$ symmetries of the LLM metric, both of which
 exchange black and white regions. One of them which can be interpreted as the particle-hole
 symmetry is the symmetry of the whole supergravity solution while the second one is just
 the symmetry of the metric and changes the sign of the fivefrom flux.
In  the second part of the paper we use  closed string probes and
their semi-classical analysis to compare the two 1/2 BPS deformations of
$AdS_5\times S^5$, the smooth LLM geometry which contains localized giant gravitons and the superstar case which is
a solution with naked singularity corresponding to smeared giants. We discuss the realization of the $\mathbb{Z}_2$ symmetry in the 
semi-classical closed string probes point of view.
 
 \end{abstract}

 \end{titlepage}

\section{Introduction}

The AdS/CFT correspondence (duality)\cite{AdS/CFT}, as the only
explicit example in which the holographic formulation of quantum
gravity (string theory) is  realized,  has
been the corner stone of the string theory studies in the past seven years. The above
correspondence is a strong/weak duality in the sense that when the
gauge theory is perturbative the string theory sigma model is
strongly coupled and vice-versa. Hence usually one can only
perform perturbative computations in one side of the duality.
Recently, however, it has been shown that there is a certain large
quantum numbers limit, the BMN sector \cite{BMN} and the
``semi-classical'' strings \cite{Gubser:2002tv,Frolov:2002av}, in
which the gauge theory and the string theory sides are both
perturbatively accessible, for a review e.g. see
\cite{BMN-review,Tseytlin}.

Besides the large quantum numbers limit, noting  the large
supersymmetry of either sides of the duality, namely \psutwo\ superalgebra,
one can focus on the BPS information which are protected by supersymmetry.
One might perform computations with BPS objects in a weakly coupled
regime on either sides of the duality and due to supersymmetry expect the
same computation to be still valid at strong coupling.
The \psutwo\ is a superalgebra with 32 supercharges and hence we have the
option of looking at various BPS sectors. The (dynamically) simplest BPS
sector is the 1/2 BPS one. In the $\N=4$ $U(N)$
gauge theory side the highest weight
state of the 1/2 BPS multiplet is a chiral primary operator. These
operators are only a function of one of the three complex scalars present
in the $\N=4$ vector multiplet. Let us denote this scalar, which is
an $N\by N$ matrix, by $Z$.
The 1/2 BPS sector, i.e. the chiral primary operators, are then all
possible gauge invariant combinations of various powers of $Z$ (and not
${Z}^\dagger$).
Since we are dealing with $N\by N$ matrices, $Z^k$ for $k>N$ are not
independent matrices.
The scaling dimension, $\Delta$, of 1/2 BPS operators is protected by
supersymmetry and is equal to their classical (engineering) dimension  which is also equal to
their $R$-charge, $J$. All the $n$-point functions of these BPS operators
are protected against gauge theory loop corrections, i.e. they are $g_{YM}$
independent. This makes 1/2 BPS sector a perfect laboratory for studying the
combinatorics and the $1/N$ expansion behavior.

Giving the $R$-charge $J$ does not completely specify the chiral primary
operator, e.g. the \opt can be single, double or multi trace and in general
a linear combination of multi-trace operators. To obtain a gauge invariant \opt made
out of product of $J$ $Z_{ij}$'s we need a $U(N)$ tensor to contract
the $N\by N$ indices, explicitly
${\cal O}_{T}=T^{i_1 i_2\cdots i_J}_{j_1 j_2\cdots j_J}\
Z_{i_1 j_1}Z_{i_2 j_2}\cdots Z_{i_J j_J}$.
Therefore, an \opt is completely specified by giving $T^{i_1 i_2\cdots i_J}_{j_1 j_2\cdots j_J}$.
This can be done through  a Young tableau made out of $J$ boxes.

It has been shown that the sector of the $\N=4$ $U(N)$ SYM on
$R\times S^3$ which is only made out of $Z$ and ${Z}^\dagger$ that
are constant on the $S^3$, is equivalent to an $N$ fermion system
in  two dimensional harmonic oscillator potential \cite{Antal,
Berenstein, Caldarelli:2004ig}. Furthermore, the 1/2 BPS sector is equivalent to
reducing the 2d fermion system to a one dimensional one. The phase
space of the $N$ fermion system can be described by the same Young
tableau discussed above \cite{Antal, Berenstein}.

In \cite{LLM} Lin-Lunin-Maldacena (LLM) constructed
the type IIB supergravity solutions which are dual
to 1/2 BPS states of ${\cal N}=4$ supersymmetric Yang-Mills
theory. As we will briefly review in section 2, these solutions are
specified by a single function which essentially obeys a six dimensional
Laplace equation, sourced on a two dimensional plane. This plane,
via AdS/CFT, should be identified with the phase space of the above mentioned
one dimensional fermion system.

The paper is organized as follows. In section 2 we will review LLM
construction and  discuss the discrete symmetries of the
solutions. We will show that there are two $\mathbb{Z}_2$
symmetries of the LLM metrics, both of which exchange black and
white regions. We study these  $\mathbb{Z}_2$ symmetries from the
supergravity, superalgebra and the ${\cal N}=4$ SYM viewpoints. In
section 3 we consider the deformation of AdS solution by adding
giant gravitons. There are two ways to do that, one leads to a
singular and the other to smooth geometries. We will probe these
geometries by closed spinning strings and discuss how the two
smooth and singular 1/2 BPS geometries are viewed by the closed
string probes. The last section is devoted to concluding remarks.

\section{$\mathbb{Z}_2$ symmetries of LLM geometries}

The AdS/CFT correspondence implies that the
type IIB supergravity solutions corresponding to
the chiral primaries of the $\N=4$ SYM should have the following properties:

{\it i)} The scaling dimension of  the 1/2 BPS operators is protected by
\susy and hence the corresponding SUGRA solution should be a static solution
with a globally defined time-like Killing vector field.

{\it ii)} The chiral primary operators are invariant under $SO(4)\subset SU(4)_R$
as well as $SO(4)\subset SO(4,2)$, therefore the SUGRA solution should have
an $SO(4)\by SO(4)$ isometry.

This class of solutions does not have non-compact
isometries other than the non-compact $U(1)$ factor associated with the translation
along the time-like Killing vector. The \susy of the system in ensured by
checking the Killing spinor equations and that they have 16 independent
solutions. In fact LLM constructed their solutions using the Killing spinors and
spinor bi-linear techniques developed in \cite{bilinears}.

The LLM solutions  are all deformations\footnote{ Classically and
just by \sugra considerations these are continuous deformations of
AdS or the plane-wave geometries. Considering the semi-classical
or quantum arguments, these deformations are, however,
parameterized by a discrete parameter \cite{LLM}. This point will
be discussed further later on.} of the two maximally \susy
solutions of type IIB backgrounds, namely $AdS_5\times S^5$ and
the plane-wave \cite{BMN-review}.\footnote{ It is worth noting
that LLM solutions are not connected to the third and the only
remaining maximally supersymmetric type IIB background
\cite{Maximal-SUSY}, the flat space.} The \susy of the LLM
geometries are then a subgroup of the AdS or plane-wave
superalgebras; specifically that is \super\ \cite{BMN-review,
TGMT}, where the $U(1)$ factor corresponds to the time
translations. In terms of ${\cal N}=4$ SYM the eigenvalues of this
$U(1)$ correspond to the scaling dimension of the operators.

Being a 1/2 BPS solution with the required supersymmetry, the dilatino variation
for the LLM geometries must be fulfilled identically. This implies that
the 1/2 BPS solutions should have a constant dilaton and the NSNS twoform
should vanish. The gravitino variation and the 1/2 BPS condition restricts the
solution further to have zero axion and RR twoform and only the selfdual
fiveform field can be turned on. The LLM geometries are given by \cite{LLM}
 \bea
 ds^2&=&-h^{-2}(dt+V_idx^i)^2+h^2(dy^2+dx^idx^i)+ye^Gd\Omega_3^2
 +ye^{-G}d{\tilde\Omega}_3^2\;,\label{LLMmetric}\cr && \cr
 h^{-2}&=&2y\cosh{G}\;,\;\;\;\;\;\;z=\frac{1}{2}\tanh{G}\;,
 \cr && \cr
 y\partial_yV_i&=&\epsilon_{ij}\partial_jz\;,\;\;\;\;\;
 y(\partial_iV_j-\partial_jV_i)=2\epsilon_{ij}\partial_yz\;,\ \ i,j=1,2\ .
 \label{bubl}
 \eea
The self dual five-form field strength is also nonzero
\be
F_5=F\wedge d\Omega_3^2+{\tilde F}\wedge d{\tilde \Omega}_3^2
\ee
where
\bea
F&=&-\frac{1}{4}\bigg{[} d\left(y^2e^{2G}(dt+V)\right)
+y^3*_3d\left(\frac{z+\frac{1}{2}}{y^2}\right)\bigg{]}\cr &&\cr
{\tilde F}&=&-\frac{1}{4}\bigg{[} d\left(y^2e^{-2G}(dt+V)\right)
+y^3*_3d\left(\frac{z-\frac{1}{2}}{y^2}\right)\bigg{]}.
\eea
To fix our notation we parameterize the two spheres as follows
 \bea
 d\Omega_3^2&=&d\theta^2_1+\cos^2\theta_1(d\theta_2^2+\cos^2\theta_2d\theta^2)\cr &&\cr
 d{\tilde \Omega}_3^2&=&d\psi_1^2+\cos^2\psi_1(d\psi_2^2+\cos^2\psi_2d\psi^2)
 \eea

As we see from the above expressions, the 1/2 BPS condition is very restrictive
and the whole solution is completely specified through a single function $z$ which
satisfies the following equation
 \be
 \partial_i\partial_iz+y\partial_y(\frac{\partial_yz}{y})=0\;.
 \label{diff}
 \ee
The above equation is a six dimensional Laplace equation for
$\Phi=\frac{1}{y^2}z$. Therefore, \eqref{diff} can be easily solved once we specify
the (delta function type) sources on the right hand side of the equation.
These sources should be placed on the $y=0$ plane
which is a two dimensional plane parameterized by $(x_1,x_2)$,
explicitly, they are of the form $\frac{1}{y}\rho(x_1,x_2)\delta(y)$.

It is reasonable to assume that only non-singular supergravity
solutions should be dual to 1/2 BPS chiral primary operators of
the gauge theory. The smoothness condition implies that $z$ must
obey the boundary condition $z=\pm\frac{1}{2}$ at $y=0$ on the
$(x_1,x_2)$ 2-plane. Let us assign black and white colors to the
$z=-\frac{1}{2}$ and $z=\frac{1}{2}$ regions of the $y=0$ plane,
respectively. This is in perfect agreement with what we expect
from the 1/2 BPS operators in the dual SYM theory: in the 2d
fermion language, the black and white regions directly correspond
to fermions (black) and holes (whites) \cite{Berenstein, LLM}.
{}From this observation, however, one learns an important fact
about the $(x_1,x_2)$ plane which is not coming out of the classical
supergravity considerations alone:\footnote{The LLM solutions are completely 
described by a single function $z$. One may try using the mini-superspace quantization of the
LLM geometries. This process has been carried out in \cite{mini-super-space} and shown that it, in fact, reproduces  \eqref{NC-plane}.} In the quantized
gravity theory the $(x_1,x_2)$ plane is a noncommutative plane
with\footnote{According to the tiny graviton matrix theory
conjecture \cite{TGMT} in the fully quantized gravity/string
theory, not only the $(x_1,x_2)$ plane but also the the spheres
$S^3$ and ${\tilde S}^3$
are also ``quantized'' fuzzy spheres \cite{TGMT, Torabian}.}%
\be\label{NC-plane} [x_1,x_2]=2\pi i\ l^4_{Pl}\ . %
\ee %
(Note that in the conventions we are using, which is the same as
the one adopted in \cite{LLM}, $x_1,x_2$ and $y$ coordinates have
dimension of length$^2$.) This in particular implies that there is
a minimal area that one can probe on the $(x_1,x_2)$ plane and
that \eqref{NC-plane} defines an orientation on the $(x_1,x_2)$
plane. The noncommutativity of the $(x_1,x_2)$ plane is necessary
to ensure that we are not going to get  solutions with naked
singularities (superstars \cite{Myers:2001aq}) as a limit of the
smooth LLM geometries.

Eq.\eqref{diff} is a linear equation and hence the most general
solutions can be constructed as linear combinations of simpler
solutions, provided that the sum is also respecting the boundary
conditions. In other words, in general a solution is specified by
a configuration of black droplets in  the $(x_1,x_2)$ plane. For
further studies in this direction see e.g. \cite{Beren2,
LLM-citations}.

\begin{figure}[htb]
\begin{center}
\epsfxsize=4in\leavevmode\epsfbox{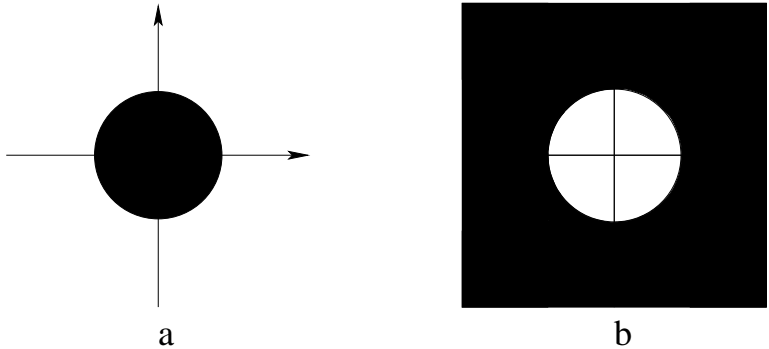}
\end{center}
\caption{The configurations which correspond to AdS$_5\times$ S$^5$. }
\label{pnfig}
\end{figure}

The simplest example is a black disc centered at the origin,
Figure \ref{pnfig} (a). Setting ${\tilde z}=z-\frac{1}{2}$ the
solution in given by%
\be\begin{split}%
{\tilde z}=&\frac{1}{2}\;\frac{r^2-r^2_0+y^2}
{\sqrt{(r^2+r_0^2+y^2)^2-4r^2r_0^2}}-\frac{1}{2}\;, \cr
V_\phi=&-\frac{1}{2}\;\frac{r^2+r^2_0+y^2}
{2\sqrt{(r^2+r_0^2+y^2)^2-4r^2r_0^2}}+\frac{1}{2},\;\;\;\;\;V_r=0\;.
 \end{split}\ee%
Here we have chosen the polar coordinates $r,\phi$ in the
$(x_1,x_2)$ plane. In terms of ${\tilde z}$ the boundary condition
is given by ${\tilde z}=-1$ for black region of radius $r_0$ and
${\tilde z}=0$ for white region and the parameter $G$ is given by%
\be%
e^{G}=\sqrt{\frac{1+{\tilde z}}{-{\tilde z}}}%
\label{GG}%
\ee%

Now let us discuss two $\mathbb{Z}_2$ symmetries of the LLM
metrics, both of which exchange black and white regions. One such
example  has been depicted in Figure \ref{pnfig}. The first
$\mathbb{Z}_2$ symmetry we would like to discuss is the one which
may be interpreted as the particle-hole symmetry in the
corresponding two dimensional fermion system \cite{Berenstein,
Ghodsi}. This $\mathbb{Z}_2$ can be discussed at supergravity
level and the LLM solutions, the superalgebra level and also in
the dual gauge theory setup and on the chiral primary operators.
This means that the $\mathbb{Z}_2$ symmetry is a non-anomalous
symmetry and remains at quantum (gravity) level.

\subsection{$\mathbb{Z}_2$ symmetry at the supergravity level}

Consider the $\mathbb{Z}_2$ transformation  generated by
\begin{subequations}\label{Z2-sym1}
\begin{align}
 G &\longleftrightarrow -G\\
(x_1,x_2)&\longleftrightarrow  (x_2,x_1)\ .
\end{align}
\end{subequations}
Under (\ref{Z2-sym1}a) it is easily seen that $z
\longleftrightarrow -z$ which is equivalent to
black$\longleftrightarrow$ white regions on the $(x_1,x_2)$ plane.
Noting \eqref{bubl}, it is easily seen that $V_idx^i$ is invariant
under the above transformation. Therefore, not only the metric,
but the whole geometry is invariant under the above $\mathbb{Z}_2$
transformation.

Eq.(\ref{Z2-sym1}a) implies that under the above symmetry we are
exchanging the two three spheres in \eqref{LLMmetric} and hence it
is exchanging the two $SO(4)$ factors of the isometries. The
(\ref{Z2-sym1}b) is nothing but the parity on the $(x_1,x_2)$
plane. This changes the orientation of the $(x_1,x_2)$ plane and
hence is not compatible with \eqref{NC-plane}.

There is another $\mathbb{Z}_2$ transformation:
\begin{subequations}\label{Z2-sym2}
\begin{align}
 G &\longleftrightarrow -G\\
t &\longleftrightarrow  -t\ ,
\end{align}
\end{subequations}
under which $V_i dx^i\longleftrightarrow -V_idx^i$, which is a
symmetry of the background metric \eqref{LLMmetric} and also
preserves \eqref{NC-plane}. This $\mathbb{Z}_2$, however, is not a
symmetry of the whole solution, because under \eqref{Z2-sym2} the
fiveform flux changes sign. Transformation \eqref{Z2-sym2}, and in
particular  (\ref{Z2-sym2}a), is again changing the black and
white regions while keeping the orientation on the $(x_1,x_2)$
plane. Therefore, there is a twofold degeneracy in the LLM
solutions for a given configuration of black and white regions on
the $(x_1,x_2)$ plane. These two solutions differ in the
orientation on the $(x_1,x_2)$ plane and/or the direction of the
time coordinate. Although the both result in the same metric, the
corresponding to fiveform fluxes differ in a sign.

\subsection{$\mathbb{Z}_2$ at the level of SUSY algebra}

The LLM geometries are 1/2 BPS and hence it is interesting to
analyze the action of the above $\mathbb{Z}_2$'s on the \susy
algebra. Before that, however, we need to introduce some fermionic
notation. The \ads\ geometry has \psutwo\ supersymmetry which is
an algebra with 32 (real) fermionic generators. The supercharges
of this algebra fall into the fermionic representations of the
$so(4,2)\simeq su(2,2)$ and $so(6)\simeq su(4)$ algebras.
Explicitly, the supercharges can be labeled as $Q_{I\hat J}$ (and
its complex conjugate) where $\hat I, J$ are Weyl indices of
$su(2,2)$ and $su(4)$, respectively. It is worth noting that for a
given  \ads\ geometry we have  freedom to choose the sign of the
corresponding fiveform flux over the $S^5$. These two \ads\
geometries, although both are maximally supersymmetric, preserve
two different \psutwo\ superalgebras. The supercharges of these
two differ in the chirality of the $su(2,2)$ and $su(4)$ fermions,
i.e. if one of them have supercharges of the form $Q_{\hat I J}$,
the other one has $Q_{\dot{\hat I}\dot J}$ supercharges ({\it cf.}
Appendix D of \cite{BMN-review}). These two \ads\ spaces come as
near horizon limit of $N$ D3-branes or anti D3-branes. Next,
consider a Weyl fermion of $su(4)$ or $su(2,2)$, $\psi_I$. This
fermion can be decomposed under $so(4)\times u(1)\subset su(4)$ or
$su(2,2)$, as $(\psi_\alpha^+,\ \psi_{\dot\alpha}^-)$, where
$\alpha,\ \dot\alpha=1,2$ are the standard four dimensional Weyl
indices and the $+$ and $-$ correspond to the $u(1)$ charge. (The
$\psi_{\dot I}$ would then decompose into $(\psi_\alpha^-,\
\psi_{\dot\alpha}^+)$.) For more detailed discussion on similar
fermionic notation see Appendix D of \cite{BMN-review}.

The
supersymmetry of the LLM solutions is then a subalgebra of either
of the two \psutwo\ superalgebras discussed above. The
supercharges $Q_{\hat I J}$ (or $Q_{\dot{\hat I}\dot J}$) under
$SO(4)\times SO(4)\times U(1)_t\times U(1)$ decompose into four
fermions which differ in the relative sign of the $U(1)$ charges.
Physically the $U(1)_t$ corresponds to the translation along the
time-like Killing direction and the charge of the other $U(1)$ is
correlated with the orientation on the $(x_1,x_2)$ plane.

The supercharges of the \psutwo , $Q_{I\hat J}$, decompose into
$(q_{\alpha\beta}^{++},\ q_{\dot\alpha\dot\beta}^{--})$ and
$(q_{\alpha\dot\beta}^{+-},\ q_{\dot\alpha\beta}^{-+})$, each set
giving rise to a \super\ algebra, which is the \susy of the LLM
geometries.
 Under $G\longleftrightarrow -G$, the $\alpha$
and $\beta$ indices are exchanged. Change of orientation on
$(x_1,x_2)$ plane (while keeping the six dimensional chirality)
implies that $\psi_{\beta}^+\longleftrightarrow
\psi_{\dot\beta}^-$. Altogether the $\mathbb{Z}_2$ symmetry
\eqref{Z2-sym1} on the fermionic indices act as
\be\label{Z2-sym1-fermion}%
\psi_{\alpha\dot\beta}^{+-}\longleftrightarrow
\psi_{\beta\dot\alpha}^{-+}\ ,\ \ \ \
\psi_{\dot\alpha\dot\beta}^{--}\longleftrightarrow
\psi_{\beta\alpha}^{++}\ .%
 \ee%
At the level of the \super\ superalgebra the $\mathbb{Z}_2$
exchanges the two $PSU(2|2)$ factors. This is exactly the same as
the $\mathbb{Z}_2$ symmetry of the plane-wave background, for an
extensive discussion on this see \cite{BMN-review}.

Therefore, if we start with a \super\ with supercharges of the
form $(q_{\alpha\beta}^{++},\ q_{\dot\alpha\dot\beta}^{--})$ we
obtain the same \super\ superalgebra where the two $PSU(2|2)$
factors are exchanged.  In other words, the $\mathbb{Z}_2$
\eqref{Z2-sym1} keeps the supergravity solution and the form of
the corresponding superalgebra invariant; similarly if we choose
to work with supercharges of the form $(q_{\alpha\dot\beta}^{+-},\
q_{\dot\alpha\beta}^{-+})$.

The second $\mathbb{Z}_2$, \eqref{Z2-sym2}, changes the fermions
as
\be\label{Z2-sym2-fermion}%
\psi_{\alpha\beta}^{++}\longleftrightarrow
\psi_{\beta\dot\alpha}^{-+}\ ,\ \ \ \
\psi_{\dot\alpha\dot\beta}^{--}\longleftrightarrow
\psi_{\dot\beta\alpha}^{+-}\ ,%
 \ee%
and hence the \super\ superalgebra is not invariant under this
$\mathbb{Z}_2$, but it goes over to another \super\ which is a
subgroup of a $PSU(2,2|4)$ algebra whose supercharges are of the
form $(q_{\alpha\beta}^{--},\ q_{\dot\alpha\dot\beta}^{++})$ and
$(q_{\dot\alpha\beta}^{+-},\ q_{\alpha\dot\beta}^{-+})$. This
\super\ is a subgroup of the $PSU(2,2|4)$ superalgebra whose
generators are in the $Q_{\dot {\hat I}\dot{ J}}$ representation.

\subsection{$\mathbb{Z}_2$ at the level of ${\cal N}=4$ SYM}

The LLM geometries are, via AdS/CFT, dual to deformations of the
${\cal N}=4$ $U(N)$ SYM by 1/2 BPS chiral primary operators. These
operators can conveniently be described by $U(N)$ Young tableaux
\cite{Antal, Torabian, Beren2} which are in turn in one-to-one
correspondence with the black and white rings on the $(x_1,x_2)$
plane \cite{LLM, Ghodsi}.

One may now ask how does the $\mathbb{Z}_2$ symmetries act on the
set of chiral primary operators or Young tableaux. {}From the
${\cal N}=4$ SYM viewpoint, there are two kinds of 1/2 BPS chiral
primary operators, they either have $\Delta=J$ or $\Delta=-J$,
where $\Delta$ and $J$ are scaling dimension and $R$-charge
respectively. The orientation in the $(x_1,x_2)$ plane is
determining the sign of $\Delta-J$. (To see this note equations
(2.24-26) of \cite{LLM} and recall that
$J=-i\frac{\partial}{\partial \phi}$ and
$\Delta=i\frac{\partial}{\partial t}$.) Under the (\ref{Z2-sym1}b)
$\mathbb{Z}_2$ then $\Delta-J\leftrightarrow -(\Delta-J)$ while
keeping $\Delta$ fixed. As we see if we choose to work with chiral
primaries made only out of $Z$, they all have $\Delta-J=0$ and
hence invariant under the orientation change in $(x_1,x_2)$ plane.
Equation (\ref{Z2-sym1}a), however, affects how the $U(N)$ indices
of the gauge theory operators are contracted, i.e. it is reflected
in the Young tableaux. At the level of the Young tableaux
exchanging black and white corresponds to exchanging the vertical
and horizontal axis of the tableau, or equivalently it is the
inversion under the line at $45^\circ$ on the Young diagram. To
see this recall the black and white assignment given a Young
tableau; the latter maybe found in \cite{Ghodsi}.\footnote{ In a
recent paper, \cite{Vijay}, a ``classical'' limit of a Young
tableau has been discussed. In this classical limit the edge of a
Young tableau which is a stairs-like line is replaced (or
approximated) with a one dimensional curve $y(x)$ where $x$ is the
radial direction in the $(x_1,x_2)$ plane. In their notation $z=
\frac{1}{2}\frac{y'-1}{y'+1}$ ({\it cf}. equation (111) of
\cite{Vijay}). The $\mathbb{Z}_2$ \eqref{Z2-sym1} corresponds to
taking $y'\to 1/y'$. That is, the $\mathbb{Z}_2$ dual Young
tableau is a curve whose tangent at each point makes the same
angle with $y=x$ line as the original curve.}

The second  $\mathbb{Z}_2$ at the level of the ${\cal N}=4$ SYM
acts like a time reversal, which despite of being a symmetry of
the action changes the chirality and representation of fermions
under the $R$-symmetry, essentially as given in
\eqref{Z2-sym2-fermion}.

\section{Deformation of AdS with giant gravitons}\label{section3}

\begin{figure}[hbpt]
\begin{center}
\epsfxsize=3.5in\leavevmode\epsfbox{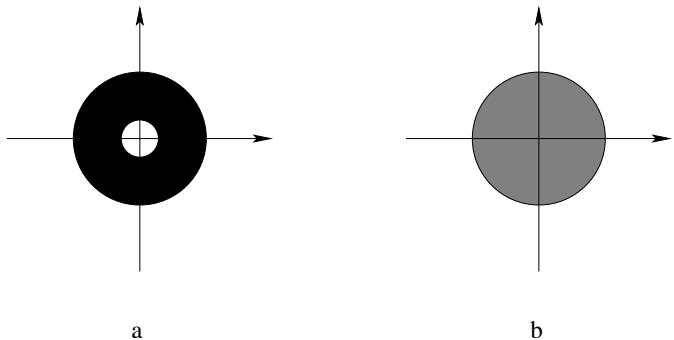}
\end{center}
\caption{$AdS$ deformation by adding giant gravitons. The deformed
solution can lead either to a smooth geometry (a) or a geometry
with naked singularity (b). In (a), $r_1$ is the radius of the
outer circle and $r_2$ the radius of the inner white circle.}
\label{star}
\end{figure}

We start with the $AdS_5\times S^5$ solution which in the LLM
construction is given by a black disk in the $(x_1,x_2)$ plane and
add spherical D3-branes (giant gravitons) wrapped on $\tilde
S^3\in S^5$. Using the LLM setup we can compute the back reaction
of the giant gravitons on the geometry. In this viewpoint the
corresponding solution is described by a black disk with a small
white hole in it. The white hole in the middle of the black disk
represents a collection of smeared  giant gravitons with maximum
size. The number of giant gravitons are fixed by the radius of the
hole (see Figure \ref{star}(a)). When the radius of the white hole
is small the good description is given in terms of the giant
gravitons, while for the large radius the better description may
be given in terms of the smeared dual giant gravitons (giants
wrapping the $S^3$ inside $AdS_5$). The corresponding supergravity
solution is given by (\ref{bubl})
with %
\be 2{\tilde
z}=\frac{r^2-r_1^2+y^2}{\sqrt{(r^2+r_1^2+y^2)^2-4r^2r_1^2}}-
\frac{r^2-r_2^2+y^2}{\sqrt{(r^2+r_2^2+y^2)^2-4r^2r_2^2}} \ee

We note that the obtained gravity is smooth without any
singularity and horizon, though adding giant gravitons might also
lead to a singular geometry. In fact the singular background
representing smeared giant gravitons has already been studied.
These are the solutions called superstars \cite{Myers:2001aq}. The
corresponding solution in the LLM notation is given by
\eqref{bubl} with
\be%
\begin{split}
{\tilde z}=&\frac{\rho}{2}
\left(\frac{r^2-r_0^2+y^2}{\sqrt{(r^2+r_0^2+y^2)^2-4r^2r_0^2}}-1\right)\\
 V_\phi=&\frac{-\rho}{2}
\left(\frac{r^2+r_0^2+y^2}{\sqrt{(r^2+r_0^2+y^2)^2-4r^2r_0^2}}-1\right)
\end{split}
\ee%
It is straightforward to check that the above ${\tilde z}$
satisfies the six dimensional Laplace equation \eqref{diff} with
the following source at $y=0$ (on the $(x_1,x_2)$ plane):%
\be\label{superstar-z}\begin{split} %
{\tilde z}=&-\rho,\;\;\;\;\;\;\;0\leq r <
r_0,\cr {\tilde
z}=&0\;\;\;\;\;\;\;\;\;\;\;\;\;\;\;\;\;\;\;\;\;r\geq
r_0, %
\end{split}\ee%
As we can see ${\tilde z}$ now takes values other than $0,-1$ and
hence the solution is singular. For the superstar solutions
\cite{Myers:2001aq},  $-1<{\tilde z}<0$ and it has been argued
that LLM solutions with $-1<{\tilde z}<0$ have generically naked,
null singularities \cite{Ghodsi, Milanesi:2005tp, Chronology}.\footnote{The values of
$\tilde z>0$ or $\tilde z<-1$ leads to spacetimes with closed
time-like curve (CTC) pathologies \cite{Ghodsi, Milanesi:2005tp,
Chronology}. As was discussed in \cite{Antal, Beren2, Ghodsi} the
dynamics of the half BPS sector of the ${\cal N}=4$ SYM or the LLM
geometries is conveniently captured by a $2d$ quantum Hall system,
or equivalently by a $3d$ (noncommutative) Matrix Chern-Simons
theory. In terms of this effective field theory, $\tilde z>0$ or
$\tilde z<-1$ corresponds to a Chern-Simons theory with level less
than one which has problems with unitarity \cite{Ghodsi}. In other
words existence of the closed time like curves is mapped into the
non-unitarity of the effective field theory description. Here we do
not consider these pathological cases.}

The ${\tilde z}$ of \eqref{superstar-z} can be conveniently
denoted by an extension of the LLM color-coding on the $(x_1,x_2)$
plane: Assign black to the region with $\tilde z=-1$, white to
$\tilde z=0$ and {\it gray} to $-1<\tilde z<0$ (this color-coding
has been employed in \cite{Ghodsi, Vijay, Jian-dai}). In this coding the closer
$\tilde z$ to $-1$, the darker the gray color and for larger
$\tilde z$ the gray is brighter. In this color-coding the above
superstar solution is a gray droplet of radius
$r_0$ located at the center (see Figure
\ref{star}(b)).\footnote{ One may wonder if gray disk can be obtained as a limit 
of black and white rings. This is of course the case.
For example consider  
a collection of $L$ successive black and white rings of area $b$ and $w$ respectively.
Then take the $L\to \infty$ limit while keeping $bL$ and $wL$ fixed. This is only possible 
if we relax the quatization of $b$ and $w$. This will lead to a superstar solution with $\rho=b/w$.
This is a simple example of the fact that a singular solution may arise as limit of smooth a geometry. 
As mentioned, for this to happen we should relax the ``quantization'' condition on 
the area in the $(x_1,x_2)$ plane. In this viewpoint, the singularity of the 
superstar geometry is removed (or ``resolved'') by the {\it quantum effects}. We would like to 
thank Jorge Russo for discussion on this ponit.
\label{footnote6}}

The black $\leftrightarrow$ white $\mathbb{Z}_2$
symmetry can be extended to the gray color-coding noting that
under the $\mathbb{Z}_2$ symmetry  $\tilde z\leftrightarrow
-1-\tilde z$ (together with a change in the orientation of the
$(x_1,x_2)$ plane), dark gray is replaced by a faint gray.
In the case of the superstar the $\mathbb{Z}_2$ symmetry then takes
$\rho\leftrightarrow 1-\rho$. 
 The $\tilde z=-1/2$ is the self-dual point.

In sum, starting from $AdS$ with $N$ units of the fiveform flux
and deforming it with giant gravitons with a given angular
momentum ${\cal J}$, we may obtain smooth (or singular) geometries
depending on whether giants are ``localized'' (or ``smeared'').
For the above giant graviton configuration 
\be\label{N-J-giant}
{\rm Giant\ graviton:}\left\{\begin{array}{cc}
&N=\frac{1}{4\pi l_p^4}(r_1^2-r_2^2)\cr
&{\cal J}=\frac{r_2^2}{4\pi l_p^4}N
\end{array}
\right.
\ee
while for the superstar case
\be\label{N-J-superstar}
{\rm Superstar:}\left\{\begin{array}{cc}
&N=\frac{1}{4\pi l_p^4}\rho r_0^2\cr
&{\cal J}=\frac{1}{8\pi l_p^4}(1-\rho)r_0^2\  N.
\end{array}
\right.
\ee
Considering a single giant as an object with angular momentum $N$, one may define number of giants
$n$ as $n=\frac{{\cal J}}{N}$ \cite{Myers:2001aq}. For the case of the giant graviton of Figure (\ref{star}(a)), $n=\frac{1}{4\pi l_p^4} r_2^2$ is (half) of the area of the inner white region 
and for the superstar case of Figure (\ref{star}(b)) $n=\frac{1}{2}\frac{1}{4\pi l_p^4}(1-\rho) r_0^2$, the extra factor of 1/2 in the 
superstar case is arising form the averaging and the fact that in this case the giants are smeared. 
For a fixed $N$, increasing ${\cal J}$ is then equivalent to adding
more giants, in the localized case this is done by making the hole
bigger while in the superstar case it means we are making the gray disk
brighter.

To compare the effects of the deformation caused by the smeared and localized giant gravitons we use semiclassical closed strings 
probing the above two backgrounds with the same $N$ and ${\cal J}$. We consider strings 
stretched along $y$ direction and rotate along some angular
directions on $S^3$ or/and ${\tilde S}^3$ and $x_1,x_2$ are set to
zero. For this particular closed
string probes the $\tilde{z}$ function is given by%
\begin{subequations}\label{ttt}
\begin{align}
{\rm AdS}:&\;\;\;\;{\tilde z}=-\frac{r_0^2}{y^2+r_0^2}\\  {\rm
Superstar}:&\;\;\;\;{\tilde z}=- \frac{\rho r_0^2}{y^2+r_0^2}\\
 {\rm Giant}:&\;\;\;\;{\tilde z}=-\frac{r_1^2}{y^2+r_1^2}+
\frac{r_2^2}{y^2+r_2^2}
\end{align}
\end{subequations}

\subsection{Closed string probes rotating in ${\tilde S}^3$}
\label{section3.1}

 For simplicity let us first consider semiclassical closed
string solutions which are stretched along $y$ direction and
rotate along one direction in ${\tilde S}^3$. The corresponding
solution and
conserved charges are given by%
\be\label{classical-rotating-string}%
 t=\kappa
\tau,\;\;\;\;\;\;\psi=\nu\tau,\;\;\;\;\;y=y(\sigma)=y(\sigma+2\pi)
\ee%
\be%
 E=\frac{\kappa}{2\pi\alpha'}\int_0^{2\pi}d\sigma \;2y\cosh G,\;\;\;\;\;
 J=\frac{\nu}{2\pi\alpha'}\int_0^{2\pi} d\sigma\;ye^{-G}
\ee%
We would now like to find the dependence of $E$ on the spin (or
R-charge) $J$. Following  \cite{Gubser:2002tv} we  consider
the short and long string limits. From  the Virasoro constraints
we get%
\begin{subequations}\label{Virasoro}
\begin{align}
{\rm AdS} :&\
{y'}^2+y^2(y^2+r_0^2)\left(\frac{\nu^2-\kappa^2}{y^2}-
\frac{\kappa^2}{r_0^2}\right)=0 \\
{\rm Superstar} :&\
{y'}^2+y^2(y^2+r_0^2)\left(\frac{\nu^2-\kappa^2}{y^2+(1-\rho)r_0^2}-
\frac{\kappa^2}{\rho r_0^2}\right)=0\\
{\rm Giant} :&\
{y'}^2+y^2(y^2+r_1^2)(y^2+r_2^2)\left(\frac{\nu^2-\kappa^2}{y^4+2y^2r_2^2+r_1^2r_2^2}-
\frac{\kappa^2}{(r_1^2-r_2^2)y^2}\right)=0
\end{align}
\end{subequations}

In general the above can be thought of as a ``zero energy'' condition
for a non-relativistic particle with a potential. Noting that the
first term, the ``kinetic'' term, is positive definite and that we
are looking for periodic solutions, the above can only be
satisfied if the ``potential'' $V(y)$ is negative or zero and in
the same  locus its derivative  is non-negative as well; that is, 
to have a periodic solution $V(y)\leq 0$ and $V'(y)\geq 0$ should
be satisfied simultaneously.

In the AdS case the potential has no minimum for any value of
$\nu,\kappa$ and hence the periodicity condition can only be
satisfied at the zeros of the potential and such zeros only exist
for $\nu\geq \kappa$. At these zeros the derivative of the
potential is always negative  except for the $\nu=\kappa$ case
where the zero of the potential is  at $y=0$. In this case we have
a string shrunk to zero size and rotating along the $\psi$
direction with the speed of light such that in the leading order
we have $E=J$. The small fluctuations of this string around the
solution \eqref{classical-rotating-string} is actually probing the
plane wave background \cite{Gubser:2002tv}.

\subsubsection{The case of superstar}

In the superstar case the situation is similarly to the AdS case,
namely the potential is always negative with negative slope.
Therefore, we won't get any closed string solution except for
$y=0$ where $V(y)$ and $V'(y)$ both vanish. This happens
independently of values of $\nu$ and $\kappa$. In the particular
case of $\nu=\kappa$, in the leading order, one finds
\be\label{E=kJ}%
E=\frac{1}{\rho}\;J %
\ee%
in which the $\rho\rightarrow 1$ is
a smooth limit that brings us back to the undeformed  in the AdS
case. 

It is interesting to note that for $\frac{1}{\rho}=k\in {\mathbb{Z}}$, $E=kJ$ is in fact similarly to the case of BPS condition $E=J$ 
but the circle along which the particle is moving
is now an $S^1/Z_k$ orbifold. Explicitly, consider $AdS_5\times S^5/Z_k$ orbifold and adopt the coordinate system in such a way that 
the orbifolding is acting on an $S^1\in S^5$. 
For the untwisted sector of the orbifold $E=kJ$ (for example see \cite{M-Sh}).

Eq.\eqref{E=kJ} becomes more interesting recalling the correspondence (equivalence) between the quantum Hall system (QHS) and the 
1/2 BPS sector of the ${\cal N}=4$ super Yang-Mills 
\cite{Beren2, Ghodsi}, according which $\rho$ is indeed the (average) density of the 
$2d$ fermions in the Landau levels (in units of the external magnetic field) \cite{Ghodsi}.
In other words, $\rho$ is equal to the filling factor. For the integer quantum Hall effect
(IQHE) $\rho$ is equal to one. For the fractional quantum Hall effect (FQHE) which is described by Laughlin wavefunctions, however,  
inverse of the filling factor is quantized in integer steps and hence for this case $\rho=1/k,\ k\in \mathbb{Z}$. \footnote{Using 
\eqref{N-J-superstar} it is readily seen that $\rho$ in terms of characteristics of the background is given by
\[ \rho=\frac{1}{1+\frac{2{\cal J}}{N^2}}\]
and hence integer $1/\rho$ happens when $\frac{2{\cal J}}{N^2}=\frac{2n}{N}\in\mathbb{Z}$.
If $n/N\in \mathbb{Z}$ then $1/\rho$ is an {\it odd} integer, corresponding to a {\it fermionic} Laughlin wavefunction for which 
inverse of the filling factor is  an 
odd number.}
That is, a superstar system  is equivalent to a fractional QHS which, as discussed  above, 
is related to an orbifolded ${\cal N}=4$ SYM, or a noncommutative Chern-Simons (NCCS) theory on 
${\mathbb{R}}\times \mathbb{R}^2/Z_k$ orbifold. This is in line with the SYM/NCCS correspondence
at levels not equal to identity proposed and discussed in \cite{Ghodsi}. This orbifold picture can be directly connected with the black/gray color-coding introduced earlier. 
Consider a black disk, perform the $Z_k$ orbifolding and again redefine the angular coordinate to cover the $(0,2\pi)$ region. In this redefinition the black region becomes $1/k$ times ``fainter'', becoming ``gray''.

One  may wonder whether the Penrose limit of this geometry
is what is seen by the small fluctuations around this zero size
string solution. We will come back to this point later when we 
study the Penrose limit of the solutions.

\subsubsection{The giant graviton case}

\begin{figure}[htb]
\begin{center}
\epsfxsize=1.5in\leavevmode\epsfbox{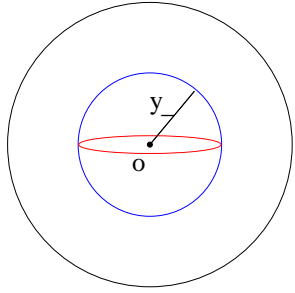}
\end{center}
\caption{Semi-classical closed string solutions stretched along $y$ direction with
angular momentum in the ${\tilde S}^3$ in the giant graviton background}
\label{GIJfig}
\end{figure}

In the giant graviton case still we do not get closed string
solution for $\nu\leq \kappa$ while for $\nu>\kappa$ there is a
possibility to get semi-classical rotating closed string solution.
The periodicity condition can be satisfied if%
\be\label{condition}%
\frac{r_1}{r_2}\geq\frac{\nu^2+\kappa^2}{\nu^2-\kappa^2} %
\ee%
for which the potential  takes positive values for $y_-\leq
y\leq y_+$, where%
\be
y_\pm^2+r_2^2=\frac{(r_1^2-r_2^2)(\nu^2-\kappa^2)}{2\kappa^2}\left(1\pm
\sqrt{1-\frac{4\kappa^2\nu^2r_2^2}{(r_1^2-r_2^2)(\nu^2-\kappa^2)^2}}\right).
\ee%
As the potential has no minimum, the only acceptable solution is
where the potential vanishes and has a positive slope at that
point. That is at $y=y_-$. An interesting feature of this case is
that the closed string cannot be longer than a maximum size given
by $ \sqrt{r_1r_2}$ which corresponds to the length of string
whose quantum numbers satisfy the equality in equation
(\ref{condition}).  (See Figure \ref{GIJfig}).

In the $r_1\gg r_2$ case one expects the effects of the giant
graviton back reaction on the geometry to be small.  To see this,
note that in this limit%
\be\label{leading-turning}%
 y_-^2\simeq \frac{\kappa^2}{\nu^2-\kappa^2}\;r_2^2. %
\ee%
Expanding (\ref{Virasoro}c) around $y_-$, in the leading order we
obtain%
\be%
 {y'}^2+y^2(y^2+r_2^2)\left(\frac{\kappa^2}{y_-^2}
-\frac{\kappa^2}{y^2}\right)=0
\ee%
which is equivalent to a semi-classical rotating closed string in
the AdS background that rotates in the AdS space with the speed
$\nu$. Of course, noting the Figures \ref{pnfig} and \ref{star},  
this is expected from the $\mathbb{Z}_2$ symmetry we have studied in the previous section 
which exchanges
the white and black regions.

We note, however, that the string can indeed distinguish this
background from pure AdS through the next-to-leading order
corrections. For example as we discussed the effects of such
contributions lead to an upper limit on the length of the longest
closed string one might have. In the leading order the turning
point is given by \eqref{leading-turning} and in the long string
limit where $\kappa\rightarrow \nu$  we can use the AdS
approximation as long as $\frac{\kappa}{\nu}<1-\frac{r_2}{r_1}$
where the string probes an AdS background with radius $r_2$ which is 
produced by the giant gravitons and the energy is given by%
\be%
E\approx J+\frac{r_2}{\pi\alpha'}\ln\frac{\alpha'J}{r_2}%
\ee%

On the other hand in the limit $\frac{\kappa}{\nu}\rightarrow
(1-\frac{r_2}{r_1})$ or $y_-\rightarrow \sqrt{r_1r_2}$ the above
approximation breaks down and the string sees the whole geometry
produced by the giant gravitons on the AdS background with the
radius $r_1$.

In the short string limit where $\nu\gg \kappa$ one may also use
the AdS approximation to find the behavior of the energy in terms
of the angular momentum $J$. In this limit we get the Regge
trajectory as in the flat space, as expected. In the short string
limit the string shrinks to zero size as%
\be%
y_-\approx \frac{r_2^2}{\sqrt{r_1^2-r_2^2}}\;\frac{\kappa}{\nu}
\ee%
and the prefactor in the Regge trajectory is changed as follows%
\be%
E^2\approx \frac{r_1^2}{\sqrt{r_1^2-r_2^2}}\;\frac{J}{\alpha'}.
\ee%
As we see, although in this case string still probes a flat space,
the slope which is the effective string tension depends on the whole geometry including the giant gravitons effects.

As a conclusion we note that as long as these closed strings
are concerned the background is very similar to AdS geometry, 
however, in the giant graviton case the folded string cannot be longer than
$L=\sqrt{r_1r_2}$. In the $r_1\gg r_2$, $L^4=n R^4$, where $n$ is the number of giants and $R$ is the AdS radius, and for large $R$ (or 
$r_1$) we can get the string as long as we want and the background is exactly
the AdS geometry given by a white whole in a black plane. From 
the previous section we note that this is also AdS solution. Our 
observation will also be supported in the next subsection by noting that
for the region smaller than $\sqrt{r_1r_2}$ the only possible solution
will be a point like string which leads to the Penrose limit of the 
geometry that would be a plane-wave, in the LLM $(x_1,x_2)$ plane notation is described by black in upper half plane and
white in lower half plane.

\subsection{Closed string probes rotating in $S^3$}

Let us now consider a semi-classical closed string solution which
is stretched along the $y$ coordinate and rotates along $\theta\in
S^3$ with speed of $\omega$. The corresponding ansatz and
conserved charges are given by%
\be%
 t=\kappa
\tau,\;\;\;\;\;\;\theta=\omega\tau,\;\;\;\;\;y=y(\sigma)=y(\sigma+2\pi)
\ee%
\be%
 E=\frac{\kappa}{2\pi\alpha'}\int_0^{2\pi}d\sigma \;2y\cosh G,\;\;\;\;\;
 S=\frac{\omega}{2\pi\alpha'}\int_0^{2\pi} d\sigma\;ye^{G}
 \label{ES}
\ee%
 In this case the Virasoro constraints read
\begin{subequations}
\begin{align}
{\rm AdS}:&\
{y'}^2+y^2(y^2+r_0^2)\left(\frac{\omega^2-\kappa^2}{r_0^2}-\frac{\kappa^2}{y^2}\right)=0\\
{\rm Superstar}:&\
{y'}^2+y^2(y^2+r_0^2)\left(\frac{\omega^2-\kappa^2}{\rho r_0^2}
-\frac{\kappa^2}{y^2+(1-\rho) r_0^2}\right)=0\\
{\rm Giant}:&\
{y'}^2+y^2(y^2+r_1^2)(y^2+r_2^2)\left(\frac{\omega^2-\kappa^2}{(r_1^2-r_2^2)y^2}
-\frac{\kappa^2}{y^4+2y^2r_2^2+r_1^2r_2^2}\right)=0
\end{align}
\end{subequations}

\begin{figure}[htb]
\begin{center}
\epsfxsize=4.5in\leavevmode\epsfbox{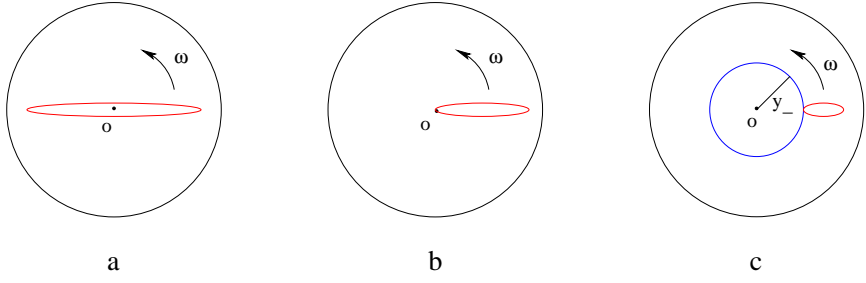}
\end{center}
\caption{Semi-classical closed string solutions stretched along
$y$ direction with angular momentum in the ``{\it AdS}'' part and
probing (a) AdS, (b) Superstar and (c) Giant graviton background.}
\label{ASGfig}
\end{figure}

In the AdS case the periodicity condition can be  satisfied for
$\omega
>\kappa$ and the turning point is given by
$y^2_0=\frac{\kappa^2}{\omega^2-\kappa^2}r_0^2$. In the long
string limit where $\omega\sim \kappa$ we get logarithmic
correction to the energy, $E\sim
S+\frac{r_0}{\pi\alpha'}\ln\frac{\alpha'S}{r_0}$, while in the
short string limit where $\omega \gg \kappa$ we find the Regge
trajectory as in the flat space $E^2\sim r_0\frac{2S}{\alpha'}$.
(See Figure \ref{ASGfig} (a)).

\subsubsection{The case of superstar}

The negative ``potential'' condition can only be satisfied when $\omega\geq \kappa$. 
The $\omega=\kappa$ case which only has a zero size string solution will be discussed  later.
For $\omega > \kappa$ the periodicity condition can be satisfied
if $\frac{\kappa^2}{\omega^2}\geq 1-\rho$ in which
the turning points are given by%
\be%
y^2_0=r_0^2\left(\rho\frac{\omega^2}{\omega^2-\kappa^2}-1\right).%
\ee%
Note, however, that unlike the AdS case the string is not folded
symmetrically around the origin. In fact we get folded orbiting
string which starts from the origin and goes up to $y_0$ and then
folds on itself. For the $y_0=0$ case, i.e. when
$\frac{\kappa^2}{\omega^2}=1-\rho$ (Figure
\ref{ASGfig} (b)), we will have a zero size string localized at
the origin.

In the long string  $\kappa\rightarrow \omega$ limit,%
\be y_0^2\sim\frac{\rho r_0^2}{\eta}\gg \rho r_0^2,\;\;\;\;\;\kappa\sim
\frac{1}{2\pi}\ln\frac{\rho r_0^2}{\eta},
\;\;\;\;\;\omega\sim\frac{1}{2\pi}\sqrt{1+\eta}\ln\frac{\rho r_0^2}{\eta}
\ee%
where $\frac{\omega^2-\kappa^2}{\kappa^2}=\eta\to 0$. Using
(\ref{ES}) in the leading order we arrive at%
\be
\begin{split}
E\sim&\frac{r_0\sqrt{\rho}}{4\pi\alpha'}\left(\frac{1}{\eta}+\ln\frac{1}{\eta}\right)
\cr
S\sim&\frac{r_0\sqrt{\rho}}{4\pi\alpha'}\left(\frac{1}{\eta}-\ln\frac{1}{\eta}\right)
\end{split}
\ee%
 or equivalently%
\be%
E\sim S+\frac{r_0\sqrt{\rho}}{2\pi\alpha'}\ln\frac{4\pi\alpha'S}{r_0\sqrt{\rho}}.
\ee%
In terms of $N$ given in \eqref{N-J-superstar}, $E-S\propto \sqrt{N/4\pi} 
\ln (4\pi S^2/N)$, which is the same as the AdS case except for the factor  2 in the second
term that appears because in this case we are dealing with a
closed string orbiting around the origin.

On the other hand in the short string limit, $\frac{\omega^2}{\omega^2-\kappa^2}=\frac{1}{\rho}+ \xi$ with
$\xi\rightarrow 0$,  $y_0^2=\xi \rho r_0^2\ll \rho r_0^2$ and using
(\ref{ES}) in the leading order we arrive at%
\be\begin{split}
E\sim&\frac{r_0}{\pi\alpha'}\;\frac{\kappa}{\sqrt{\omega^2-\kappa^2}},\cr
S\sim&\frac{(1-\rho)r_0}{\pi\alpha'}\;\frac{\omega}{\sqrt{\omega^2-\kappa^2}}.
\end{split}\ee %
Therefore,%
\be%
E^2\sim
-\frac{r_0^2}{\pi^2{\alpha'}^2}+\left(\frac{1}{1-\rho}\right)^2 S^2,
\ee%
which is not the Regge trajectory in the flat space as expected.
We note that there is a lower limit on the spin of the string,
$S^2\geq \left(\frac{1-\rho}{\rho}\right)^2 \frac{\rho r_0^2}{\pi^2 \alpha'^2}
$ or in terms $N, {\cal J}$:\footnote{Here we have assumed $l_p^2=\alpha'$.}
\be\label{S-limit}
\frac{S^2}{N}\geq \frac{1}{\pi} \left(\frac{{\cal J}}{N^2}\right)^2
\ee
and as expected in the $\rho\to 1 ({\cal J}\to 0)$ limit the above lower bound on $S$ goes to zero.
Taking the above  bound on $S$ into account it is readily seen that
\[
E^2\geq \frac{4}{\pi}\  \frac{2{\cal J}}{N} (1+\frac{2{\cal J}}{N^2})
\]

Moreover, there is no smooth
$\rho\rightarrow 1$ limit which means that this state is not
present in the AdS background. On the
other hand for large $S$ limit we find%
\be\label{E=S(1-rho)}
 E\approx
\frac{1}{1-\rho}\;S-\frac{1}{2\pi^2{\alpha'}^2}\;\frac{(1-\rho)r_0^2}{S}.
\ee %
The above equation should be compared with \eqref{E=kJ}. As we see in the leading order in $S$
\eqref{E=S(1-rho)} is obtained from \eqref{E=kJ} by $\rho\to 1-\rho$, the $\mathbb{Z}_2$ transformation. This is in agreement with our earlier arguments about the  $\mathbb{Z}_2$
which exchanges the $S^3$ and $\tilde S^3$.
For integer values of $1/1-\rho$ \eqref{E=S(1-rho)} may correspond to an orbifold probed by closed strings ({\it cf.} discussions of section 3.1.1).

Eq.\eqref{E=S(1-rho)} has the same linear behavior as in the AdS in the leading
term.  One might then wonder if the small fluctuation of this
string probes a plane wave geometry as well. We will return to this question in the next section.

It worth noting that in the superstar case we never get Regge trajectory which 
reflects the fact that the solution is singular. Actually we would expect to get
flat Regge trajectory in the core of a solution if the effects of curvature
is negligible. But in this case the geometry is singular exactly where we would 
expect to get flat Regge trajectory and therefore  absence of the Regge trajectory signals the singularity of the background at the origin, at $x_1=x_2=y=0$.
We note also that, excluding the regime near the singularity, the other parts
seen by rotating closed strings are essentially the same as the AdS background. In this sense the
superstar geometry is closer to AdS backgraond then the giant graviton geometry.

\subsubsection{The giant graviton case}

In the giant graviton case the periodicity condition is not
satisfied for $\omega<\kappa$. While for $\omega \geq \kappa$ one
may have the closed string solution provided that%
\be%
\frac{r_1}{r_2}\geq \frac{2\omega^2-\kappa^2}{\kappa^2},
\label{rrr}
\ee%
in which the turning points of the closed string are given by%
\be
y_{\pm}^2+r_2^2=\frac{\kappa^2(r_1^2-r_2^2)}{2(\omega^2-\kappa^2)}\left(1\pm
\sqrt{1-\frac{4\omega^2r_2^2(\omega^2-\kappa^2)}{\kappa^4(r_1^2-r_2^2)}}\right).
\ee%
We note, however, that unlike the AdS case we will get folded
orbiting closed string stretched along $y$ direction for $y_-\leq
y\leq y_+$ (see Figure \ref{ASGfig} (c)). We also note  that $y_-$
changes from zero up to an upper limit given by $\sqrt{r_1r_2}$
and reaches the bound, $y_-=\sqrt{r_1r_2}$, for $\frac{r_1}{r_2}=
\frac{2\omega^2-\kappa^2}{\kappa^2}$ where we get  zero size
strings localized at $\sqrt{r_1r_2}$. For $\omega=\kappa$ only
zero size string satisfies the condition. The situation is
as follows.

Let us start from the limit where $y_-=y_+=\sqrt{r_1r_2}$ in
which the string has zero size,
localized at $\sqrt{r_1r_2}$ and with energy linearly proportional to its
spin:%
\be%
E=\frac{1}{2}(1+\frac{r_1}{r_2})S. \label{yy}
\ee%
This happens when the equality in (\ref{rrr}) has been satisfied.
We note that the $r_2\rightarrow 0$ is not a smooth limit and
therefore this is a new sector that has occurred because of the presence of giant gravitons.

One may then change situation a little bit so that
$y_\pm=\sqrt{r_1r_2}\pm \epsilon$ where we would have a short
closed string with length $2\epsilon$. For $\omega-\kappa\sim
{\cal O}(1)$ we get closed string with the length of
\[
l=\frac{\sqrt{\kappa^4(r_1^2-r_2^2)-4\omega^2r_2^2(\omega^2-\kappa^2)}}
{(\omega^2-\kappa^2)}\sqrt{\frac{r^2_1-r^2_2}{r_1r_2}}.
\]
When $\kappa$ approaches $\omega$ we will have long closed string
and  in the  $\kappa\rightarrow \omega$ limit the string is
stretched in $y$ direction between the origin, $y_-\rightarrow 0$,
and infinity $y_+\rightarrow \infty$. This might be thought  as
the case when the periodicity condition is going to be lost and we
are dealing with open string stretched all the way to infinity.
Actually as we will discuss in section 3.4 the better
description could be be given in terms of zero size string
localized at the origin whose energy is linearly dependent on the
spin at leading order, E=S, and the  background observed by the
fluctuations around this zero size classical solution would be the
plane-wave solution.

\subsection{Multi-spin string probes}

In this section we briefly study the multi-spin closed string
solutions in the backgrounds that we have been considering. In
general since the LLM backgrounds have $SO(4)\times SO(4)\times
U(1)_+$ isometry the most general solution one can consider is
labeled by five quantum numbers, the energy and four quantum spin
quantum numbers \cite{Frolov:2003qc}. To be more precise let us
consider the case with $r=0$ and $\phi={\rm constant}$ where the
Polyakov action reads
\bea%
I=-\frac{1}{4\pi\alpha'}\int d\tau d\sigma\!\!\!\!\!
&\bigg{[}&\!\! -h^{-2}\partial_\alpha t\partial^\alpha t +h^2
\partial_\alpha y\partial^\alpha y\\
&+& ye^G(\partial_\alpha \theta\partial^\alpha \theta+
\sin^2\theta
\partial_\alpha \psi_1\partial^\alpha \psi_1+ \cos^2\theta
\partial_\alpha \psi_2\partial^\alpha \psi_2) \cr
&+&ye^{-G}(\partial_\alpha \beta\partial^\alpha \beta+ \sin^2\beta
\partial_\alpha \gamma_1\partial^\alpha \gamma_1+ \cos^2\beta
\partial_\alpha \gamma_2\partial^\alpha \gamma_2)\bigg{]}\nonumber
\eea%
Note that we have changed the parametrization of two spheres to
make the isometries manifest. For the isometry of the theory we
have the following conserved charges%
\begin{subequations}
\begin{align}
E=\frac{1}{2\pi\alpha'}\int & d\sigma h^{-2}\partial_\tau t,\\
S_1=\frac{1}{2\pi\alpha'}\int d\sigma\; ye^G\sin^2\theta\;
\partial_\tau\psi_1, & \;\;\;\;\;\; S_2=\frac{1}{2\pi\alpha'}\int
d\sigma\; ye^G\cos^2\theta\;
\partial_\tau\psi_2,\\
J_1=\frac{1}{2\pi\alpha'}\int\; d\sigma ye^{-G}\sin^2\beta\;
\partial_\tau\gamma_1, & \;\;\;\;\;\; J_2=\frac{1}{2\pi\alpha'}\int\;
d\sigma ye^{-G}\cos^2\beta\;
\partial_\tau\gamma_2.%
\end{align}
\end{subequations}

\subsubsection{The $S_2=J_2=0$ case}
As the first multi-spin example we consider the case in which the
string
rotates both in $S^3$ and ${\tilde S}^3$ as%
\be%
t=\kappa \tau,\;\;\;\;\;
\theta=\omega\tau,\;\;\;\;\;\psi=\nu\tau,\;\;\;\;\;y=y(\sigma)=y(\sigma+2\pi)
\label{SOL} \ee %
where the Virasoro constraints lead to%
\begin{subequations}\label{nu-kappa-omega}%
\begin{align}
{\rm AdS}:&\
{y'}^2+y^2(y^2+r_0^2)\left(\frac{\omega^2-\kappa^2}{r_0^2}+\frac{\nu^2-\kappa^2}{y^2}\right)=0\\
{\rm Superstar}:&\
{y'}^2+y^2(y^2+r_0^2)\left(\frac{\omega^2-\kappa^2}{\rho r_0^2}
+\frac{\nu^2-\kappa^2}{y^2+(1-\rho)r_0^2}\right)=0\\
{\rm Giant}:&\
{y'}^2+y^2(y^2+r_1^2)(y^2+r_2^2)\left(\frac{\omega^2-\kappa^2}{(r_1^2-r_2^2)y^2}
+\frac{\nu^2-\kappa^2}{y^4+2y^2r_2^2+r_1^2r_2^2}\right)=0
\end{align}
\end{subequations}%
 The general features of this solution is the same as
what we have studied before. Indeed for $\nu<\kappa<\omega$ the
physics is the same as the case  the string rotates only in the
``AdS'' part (has only $\omega$) while for $\omega<\kappa<\nu$ the
string as if  only $\nu\neq 0$ as we discussed above.

\subsubsection{The $J_1=J_2=0$ case}

 As the next example let us consider two-spin solution in ``AdS''
part of the metric. That is, we are looking for a solution
describing a closed string rotating in both $\psi_1$ and $\psi_2$
where the
ansatz would be%
\be t=\kappa
\tau,\;\;\psi_1=\omega_1\tau,\;\;\psi_2=\omega_2\tau,\;\;y=y(\sigma)=y(\sigma+2\pi),
\;\;\theta=\theta(\sigma)=\theta(\sigma+2\pi).%
\ee%
The equations of motion and the Virasoro constraints are then given
by%
\bea
&&(h^2y')'=\partial_y(h^2){y'}^2+\partial_y(h^{-2})\kappa^2+\partial_y(ye^{G})({\theta'}^2-
\omega_1^2\sin^2\theta-\omega_2^2\cos^2\theta),\cr
&&(ye^{G}\theta')'=\frac{1}{2}(\omega_2^2-\omega_1^2)ye^{G}\sin2\theta,\\
&&h^2{y'}^2+ye^{G}{\theta'}^2=h^{-2}\kappa^2-ye^{G}(\omega_1^2\sin^2\theta+\omega_2^2\cos^2\theta)
\nonumber \eea%

It is of course very difficult to solve these system of non-linear
equations for generic values of parameters, though, it can be
simplified by setting $\omega_1=\omega_2=\omega$ which implies%
\be%
\theta'=\frac{c}{ye^{G}}%
\ee%
for a constant $c$. For zero $c$ we recover the single spin case
discussed in  the previous sections,
while for nonzero $c$ the Virasoro constraints reads as%
\begin{subequations}
\begin{align}
{\rm AdS}:&\
{y'}^2+y^2(y^2+r_0^2)\left(\frac{\omega^2-\kappa^2}{r_0^2}+\frac{c^2/y^2-\kappa^2}{y^2}\right)=0\\
{\rm Superstar}:&\
{y'}^2+y^2(y^2+r_0^2)\left(\frac{\omega^2-\kappa^2}{\rho r_0^2}
+\frac{c^2/y^2-\kappa^2}{y^2+(1-\rho) r_0^2}\right)=0\\ {\rm Giant}:&\
{y'}^2+y^2(y^2+r_1^2)(y^2+r_2^2)\left(\frac{\omega^2-\kappa^2}{(r_1^2-r_2^2)y^2}
+\frac{c^2/y^2-\kappa^2}{y^4+2y^2r_2^2+r_1^2r_2^2}\right)=0\label{multi-spin-giant}
\end{align}\end{subequations}%
One can now proceed with the analysis of this equation to check
whether the periodicity condition is satisfied and we get closed
string solutions. We can then see how the string  probe views
different backgrounds. This multi-spin solution for AdS case has
been studied in \cite{Frolov:2003qc}. In the short string limit
where the closed string is near the center of AdS we get usual
Regge trajectory as in the flat space plus a correction due to the
curvature of AdS. In the long string case where the string is close to the
boundary of AdS one finds
\be
E\approx 2S+\frac{3}{4}\frac{r_0^{2/3}}{{\alpha'}^{2/3}}\; (4S)^{1/3}+{\cal O}(S^{-1/3})
\ee
which shows the first correction to $E-2S$ goes as $S^{1/3}$ which is different from
the logarithmic correction in the single spin case. Of course we note that this solution with 
large $S$ is not stable \cite{Frolov:2003qc}.

In the remaining part of this subsection we will only briefly consider the giant 
graviton and superstar cases and postpone the detail to the future studies.
Viewing \eqref{multi-spin-giant} as the zero energy condition with
a given potential ({\it cf}. discussions of section
\ref{section3.1}), the periodicity condition can be satisfied only
when the potential is negative or zero and has a negative slope in
the same region. This can be achieved when
\be%
4(\omega^2r_2^2+c^2)(\omega^2-\kappa^2)\leq
\kappa^4(r_1^2-r_2^2),%
\ee%
in which the turning points are given by%
\be%
y^2_\pm+r_2^2=\frac{\kappa^2(r_1^2-r_2^2)}{2(\omega^2-\kappa^2)}\left(
1\pm\sqrt{1-\frac{4(\omega^2r_2^2+c^2)(\omega^2-\kappa^2)}{\kappa^2(r_1^2-r_2^2)}
}\right).
\ee%
Here we just consider the simplest example where $y=y_0$ is constant
which means that the
string becomes circular and is stretched only in
$\theta$ direction. For this situation we get%
\be%
{\theta'}^2=f_1\kappa^2-\omega^2,\;\;\;\;\;{\theta'}^2=-f_2\kappa^2+\omega^2\ ,
\ee%
where%
\be%
f_1=\frac{(y_0^2+r_1^2)(y_0^2+r_2^2)}{y_0^4+2y_0^2r_2^2+r_1^2r_2^2},\;\;\;\;\;
f_2=\frac{y_0^2(y_0^2+r_2^2)^2+r_2^2(y_0^2+r_1^2)^2}{(y_0^2+r_2^2)(y_0^4+2y_0^2r_2^2+r_1^2r_2^2)}\ .
\ee%
One can easily solve this equation which gives $\theta=w\sigma$
where $w$ is the winding number of the string around $\theta$. For
the $w=1$ case we get
\be%
\kappa^2=2\frac{(y_0^2+r_2^2)(y_0^4+2y_0^2r_2^2+r_1^2r_2^2)}{(y_0^4-r_2^2r_1^2)(r_1^2-r_2^2)},\;\;\;\;\;
\omega^2=\frac{(2y_0^2+r_2^2+r_1^2)(y_0^4+2y_0^2r_2^2+r_1^2r_2^2)}{(y_0^4-r_2^2r_1^2)(r_1^2-r_2^2)}\ .
\label{KO}
\ee%
The conserved charges for this solution are given by%
\be%
E=\frac{\kappa}{\alpha'}h^{-2}_0,\;\;\;\;\;S_1=S_2=S=\frac{\omega}{2\alpha'}(ye^G)_0\ .%
\ee%
Note that both $E$ and $S$ are functions of $\omega$ and $\kappa$
and $y_0$. By making use of (\ref{KO}) one may find $E=E(y_0)$ and
$S=S(y_0)$ as follows \be
E=\frac{\sqrt{2}}{\alpha'}\;\frac{(y_0^2+r_1^2)(y_0^2+r_2^2)^{3/2}}{(y_0^4-r_2^2r_1^2)^{1/2}(r_1^2-r_2^2)},
\;\;\;\;\;
S=\frac{1}{2\alpha'}\frac{(2y_0^2+r_2^2+r_1^2)^{1/2}(y_0^4+2y_0^2r_2^2+r_1^2r_2^2)}
{(y_0^4-r_2^2r_1^2)^{1/2}(r_1^2-r_2^2)} \ee One can now eliminate
$y_0$ from these expression to find the dependence of energy on
the spin. For example in the large $y_0$ limit one gets%
\be
E\approx
2S+\frac{3}{4}\;\frac{(r_1^2-r_2^2)^{1/3}}{{\alpha'}^{2/3}}
(4S)^{1/3}+{\cal O}(S^{-1/3})%
\ee%
 which has the same form as in the
AdS case \cite{Frolov:2003qc}, namely the first correction to
$E-2S$ goes as $S^{1/3}$ unlike the single spin where we had
logarithmic correction.

On the other hand unlike the AdS case we cannot get small $y_0$
limit and therefore we wont get Regge trajectory for short string
limit \cite{Frolov:2003qc}. In fact from the Virasoro constraint
we observe that in order to get a well-behaved solution one needs
to have
$y_0^2=\frac{2\omega^2-\kappa^2}{\kappa^2}r_2^2+\frac{2c^2}{\kappa^2}$
or%
\be%
y_0^4=r_1^2r_2^2+\frac{4c^2}{\kappa^4}(c^2+\omega^2r_2^2)\ .
\ee%
Therefore, the shortest string one can have is of order of
$\sqrt{r_1r_2}$ which is given in the limit of $c\rightarrow 0$
where we get the single spin solution. As a conclusion the
circular multi-spin closed strings only exist for radius bigger
that $\sqrt{r_1r_2}$ where we get the same behavior as in the AdS case.

One can also do the same computations for superstar case. For the 
superstar case in the long string limit we get exactly the same result as AdS
case {\it i.e.}
\be
E\approx 2S+\frac{3}{4}\frac{\rho^{1/3} r_0^{2/3}}{{\alpha'}^{2/3}}\; (4S)^{1/3}+{\cal O}(S^{-1/3})\ .
\ee
On the
other hand in the short string case we will get
\be
E\approx (\frac{1}{1-\rho})^{1/2} S
\ee
which is not the Regge trajectory and in fact is very similar to the single spin case
where we get linear behavior. This again confirms our observation that apart
the near core limit the superstar solution behaves very similar to AdS and the effect
of the singularity  changes the behavior of the string near the core where unlike
the AdS case we won't get Regge trajectory of flat space. Instead we get a linear
behavior which has no smooth $\rho\rightarrow 1$ limit reflecting the fact that
the geometry is singular and the string feels the singularity.

\subsection{The plane-wave limits}

\begin{figure}[htb]
\begin{center}
\epsfxsize=3.5in\leavevmode\epsfbox{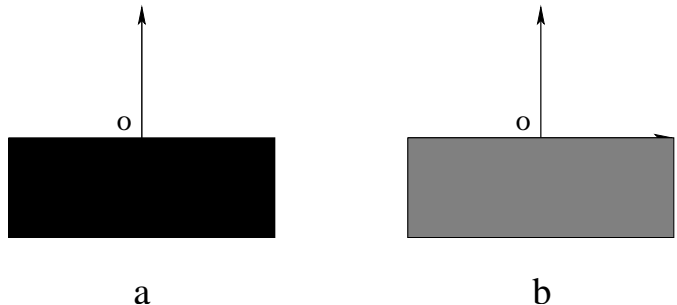}
\end{center}
\caption{The boundary conditions corresponding to the plane-wave limit of
AdS (a) and superstar (b) solutions.}
\label{PenASfig}
\end{figure}

As we have seen in the previous subsections there are cases where
the string gets zero size. We have also studied the dependence of
energy on spin in the leading order. Of course the better
treatment would be to study small fluctuations around these point
like string solutions. For example in the AdS background for the
closed string which rotates only in $S^5$ part the periodicity
condition can only be satisfied if the string is shrunk to zero size,
localized at the origin while rotating with the speed of light.
The fluctuation around this classical solution leads to the 
plane-wave solution of the AdS background
\cite{{Gubser:2002tv},{Frolov:2002av}}. From LLM point of view
this can be done by focusing on a small region around the edge of
the disk in the $(x_1,x_2)$ plane and then blowing up this region.
The boundary condition we get  corresponds to the plane-wave
solution (see Figure \ref{PenASfig}(a)).
Practically the procedure can be done by taking  the limit of
$r_0\rightarrow \infty$ and keeping $x_1, x_2$ and $w$ fixed,
where \be
r-r_0=\frac{x_2}{r_0},\;\;\;\;y=\frac{w}{r_0},\;\;\;\;\phi-\frac{\pi}{2}=\frac{x_1}{r_0^2}
\ee which leads to%
\be%
z=\frac{x_2}{2\sqrt{x_2^2+w^2}},\;\;\;\;V_1=-V_\phi\partial_1\phi=\frac{1}{2\sqrt{x_2^2+w^2}},\;\;\;\;\;V_2=0
\ee%

Having had the zero size string in the superstar and giant
graviton cases, one might be wondering whether the same physics can
appear there. It is very messy to study small
fluctuations around this classical solutions for these cases, nonetheless one
may follow the above procedure to find the plane-wave limit in
these cases as well.

In the superstar case we can again focus on a region near the edge
of the gray disk and then blow up the region. We note, however,
that there are two different ways to do that. If we focus on the
region and then blow it up by taking $r_0\rightarrow \infty$ and $\rho\rightarrow 1$ limit
while keeping $(1-\rho)r_0$  fixed one will get the  boundary
condition as in Figure \ref{PenASfig} (a), namely the geometry we
obtain is the plane-wave limit of AdS solution. On the other hand
one may consider the case where $r_0$ goes to
infinity while keeping $\rho$ fixed. More precisely
\be%
r-r_0=\frac{x_2}{r_0},\;\;\;\;\;
y=\frac{w}{r_0},\;\;\;\;\phi-\frac{\pi}{2}=\frac{x_1}{r_0^2}.
\ee%
In this limit one gets
\be%
{\tilde
z}=\frac{\rho}{2}\left(\frac{x_2}{\sqrt{x_2^2+w^2}}-1\right),\;\;\;\;\;
V_1=\frac{\rho}{2}\;\frac{1}{\sqrt{x_2^2+w^2}},\;\;\;\;\;V_2=0\ ,
\ee%
which corresponds to a singular plane-wave solution given by the
boundary condition as in Figure \ref{PenASfig} (b). Similarly to
the superstar solution this Penrose limit heads to a space with  null, naked
singularity.

In sum,  there are two ways of taking the Penrose limit of half
BPS superstar solution which has a naked singularity. One leads to
the smooth maximally supersymmetric plane-wave geometry and the
other to a 1/2 BPS solution with naked singularity. In this
regards it is quite similarly to the Penrose limits of $AdS_5\times
S^5/Z_k$ orbifolds \cite{M-Sh} (recall also the discussions of section 3.1.1).

\begin{figure}[htb]
\begin{center}
\epsfxsize=1.5in\leavevmode\epsfbox{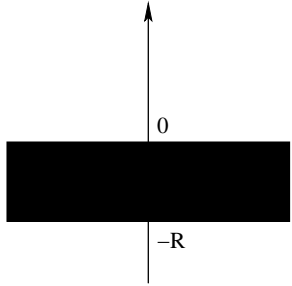}
\end{center}
\caption{Penrose limit of deformed AdS by giant gravitons.}
\label{penfig}
\end{figure}

In the case of the giant graviton similarly there are two Penrose
limits.  In the first one we blow up the edge of the droplet by
taking large $r_1$ limit while keeping $r_2$ fixed. This leads to
the plane-wave limit of AdS, as if there are no giant gravitons.
There is also another limit one may consider, namely
\cite{ Ebrahim:2005uz, Takayama:2005bc}%
\be%
y=\frac{w}{r_1},\;\;\;\;\;r-r_1=\frac{x_2}{r_1},\;\;\;\;\;\;
r_1-r_2=\frac{R}{r_1},\;\;\;\;\;
\phi-\frac{\pi}{2}=\frac{x_1}{r_1^2}%
\ee%
which in the limit of $r_1\rightarrow \infty$ keeping $w,x_1,x_2$
and $R$ fixed, we get%
\be
\begin{split}
 {\tilde
z}&=\frac{x_2}{2\sqrt{x_2^2+w^2}}-\frac{x_2+R}{2\sqrt{(x_2+R)^2+w^2}},\;\;\;\;V_2=0,\\
V_1&=\frac{1}{2\sqrt{x_2^2+w^2}}-\frac{1}{2\sqrt{(x_2+R)^2+w^2}}\ .
\end{split}
\ee%
In this case we send the size of the giant gravitons to infinity in
the same rate as $r_1$ or $r_2$ and  hence after the limit the
spherical brane becomes a flat three brane. For this solution the
boundary condition is given by a long strip in the $(x_1,x_2)$
plane as Figure \ref{penfig}.

\section{Discussions and Conclusions}

In this paper we compared two half BPS deformations of $AdS_5\times S^5$,  one is LLM type which is a smooth deformation of $AdS_5\times S^5$ with some number of giant gravitons on the
$S^5$ and the other is of the form of superstar with naked singularity. We chose the two backgrounds to have the same number of the fiveform flux over the $S^5$, $N$,  
and with the same angular momentum along an $S^1\in S^5$, ${\cal J}$ and used closed string probes.
As we showed closed strings probes can distinguish the two backgrounds and the fact that the superstar geometry is singular. In particular we did not find Regge trajectory behavior for 
strings in the superstar case, showing that the string probe feels a region with large 
curvature, rather than an approximately flat space, a sign of naked curvature singularity.

In section two of this paper we introduced and discussed a $\mathbb{Z}_2$ symmetry of the
LLM backgrounds, which can also be extended to the superstar geometries. As we discussed the
closed string probes also show the same $\mathbb{Z}_2$ symmetry. 

One of the intriguing points discussed was the relation between energy and the angular momentum and/or spin of the closed string probe for the superstar case, \eqref{E=kJ}, \eqref{E=S(1-rho)}. Of course these relations do not have any counterpart in the LLM geometry of giant gravitons.  As discussed for integer
values of $1/\rho$ or $1/1-\rho$ this relations are exactly those one would expect from a short string probing $AdS_5\times S^5/Z_k$ or $AdS_5/Z_k\times S^5$ 1/2 BPS orbifolds, where 
$1/\rho$ and $1/1-\rho$ are equal to $k$. In our case, however, we did not have any argument
as to why $1/\rho$ should be an integer. This condition may arise as a consistency condition once we quantize our semi-classical closed string probes. We also mentioned a possible  relation
to fractional quantum Hall effect. 
As we discussed the superstar solution can be obtained as a ``classical'' limit of a 
smooth LLM geometry ({\it cf.} footnote \ref{footnote6}).  One may then wonder whether the converse if also true, i.e. whether a smooth LLM type geometry can be obtained from 
a collection of superstars. Recalling the superposition property of the LLM type solutions, this may seem possible for integer values of $1/\rho$. For example one may expect to obtain $AdS$ 
geometry, a black disk, from $1/\rho$ number of superstars, gray disks, sitting on top of each
other. 
Clarifying this relation and the quantization of $\rho$
(which in the language of the corresponding quantum Hall system is nothing but the density of fermions or the filling factor) is a question with an obvious interest. This is postponed to feature works.

In this paper we have only considered the closed string solution, though one
could also study open strings as well. In this case
we should drop the periodicity condition in $\sigma$ and instead
consider an open string which would
have been string stretched all the way to boundary if we had been in the pure AdS. But in our case
one could interpret them as open strings ending on the giant gravitons. Actually one might imagine
then  as closed strings opened up  and attached on giant gravitons.

As we have seen in the giant graviton case the closed string with angular momentum in the ``AdS''
part it is not possible to be folded symmetrically around the origin and we could only find an orbiting folded
string. The situation is very similar to the case when we have horizon, like the case of AdS-Schwarzchild blackhole. We note, however, that our background solutions have no horizon.
As we see the folded closed strings can easily recognize the giant graviton background from
the pure AdS solution. 
 In the AdS case one only has closed string folded symmetrically around the origin, whereas in the giant case the string orbits
around the origin. One may think that adding giant gravitons will split the closed string into
two parts (or open them up with the ends on the giants) and therefore the limit going from giant graviton to pure AdS is not  smooth.
In particular in the giant graviton case there is string localized at $\sqrt{r_1r_2}$ which
does not have counterpart in the pure AdS case. In fact as we see from (\ref{yy}) the $r_2\rightarrow 0$ limit is not a smooth one. From closed strings point of view, it does not
matter how small $r_2$ is, as soon as it is nonzero the string probes a new physics. This
might be interpreted as the fact how the string could probe the quantum nature of the $(x_1,x_2)$ plane.

As has been discussed in the literature \cite{Milanesi:2005tp, Chronology} the superstar case
with $\rho>1$ corresponds to a geometry with closed time-like curves, a background with naked timelike singularity. It is interesting to briefly analyze this case from the closed string probes viewpoint. Let us again recall (\ref{nu-kappa-omega}b):
\[
{y'}^2+y^2(y^2+r_0^2)\left(\frac{\omega^2-\kappa^2}{\rho r_0^2}+
\frac{\nu^2-\kappa^2}{y^2+(1-\rho)r_0^2}\right)=0 %
\]%
For $\rho>1$ there are poles at
$y=\pm\sqrt{(\rho-1)}r_0$ which changes the situation drastically. 
Further study of
this case which can shed further light on the nature of these closed time-like curves
is postponed to future works.



\section*{Acknowledgements}

We would like to thank Ahmad Ghodsi, Jorge Russo, David Tong  and especially Esmaeil
Mosaffa for fruitful discussions. M. A. would like to thank CERN theory
division for very warm hospitality. H. E. would also like to thank
CERN theory division and DAMTP high energy particle physics group for 
very warm hospitality.

\end{document}